\documentclass[pra,twocolumn,nofootinbib,eqsecnum,floatfix,showpacs]{revtex4}
\usepackage{pslatex}
\usepackage{epsfig}
\usepackage{bm}% bold math
\usepackage{amsmath}
\input epsf
\numberwithin{equation}{section}

\def\etal{{\it et al.}}

\begin{document}

%\centerline{\Huge DRAFT \today}
%\vskip .3in

\title{General Relativity in the\\
Undergraduate Physics Curriculum}

\author{James B.~Hartle}

\email{hartle@physics.ucsb.edu}

\affiliation{Department of Physics\\
University of California, Santa Barbara, CA 93106-9530}

\date{\today}

\begin{abstract}

Einstein's general relativity is increasingly important in contemporary physics on the frontiers
of both the very largest distance scales (astrophysics and cosmology) and the very smallest(elementary particle physics). This paper makes the case for 
a `physics first' approach to introducing
general relativity to undergraduate physics majors.

\end{abstract}

\pacs{01.40.-d,04.20.-q,95.30.Sf}

\maketitle

\section{Introduction}

Einstein's 1915 relativistic theory of gravity --- general relativity --- will soon be a
century old.  It is the classical theory of one of the four fundamental forces.  It underlies
our contemporary understanding of the big bang, black holes, pulsars, quasars, X-ray sources,
the final destiny of stars, gravitational waves, and the evolution of the universe itself.  It
is the intellectual origin of many of the ideas at play in the quest for a unified theory of
the fundamental forces that includes gravity.  The heart of general relativity is one of the
most beautiful and revolutionary ideas in modern science --- the idea that gravity {\sl is} the
geometry of curved four-dimensional spacetime.  General relativity and  quantum
mechanics are usually regarded as the two greatest developments of twentieth-century physics.

Yet, paradoxically, general relativity --- so well established, so important for several
branches of physics, and so simple in its basic conception --- is often not represented
anywhere in the undergraduate physics curriculum. An informal survey by William Hiscock \cite
{His05} of the course offerings of 32 mid-western research universities found only a handful
that offered an intermediate (junior/senior) course in general relativity as part of the
undergraduate physics curriculum.  This has the consequence that many students see gravity
first in the context of planetary orbits in basic mechanics and next, if at all, in an advanced
graduate course designed in part for prospective specialists in the subject.  There might have
been an argument for such an organization half a century ago. But there is none today in an era
when gravitational physics is increasingly important, increasingly topical, increasingly 
integrated with other areas of physics, and increasingly connected with experiment and
observation.  In the author's opinion, every undergraduate physics major 
should have an opportunity to be introduced to general relativity.

Its importance in contemporary physics is not the only reason for introducing undergraduates to
general relativity.  There are others: First, the subject excites interest in students.  Warped
spacetime, black holes, and the big bang are the focus both of contemporary research and of
popular scientific fascination.  Students specializing in physics naturally want to know more.
A further argument for undergraduate general relativity is accessibility.  As I hope to show in
this paper, a number of important phenomena of gravitational physics can be efficiently
introduced with just a basic background in mechanics and a minimum of mathematics beyond the
usual advanced calculus tool kit.  Other subjects of great contemporary importance such as high
temperature superconductivity or gauge theories of the strong interactions require much more
prerequisite information.  General relativity can be made accessible to both students and
faculty alike at the undergraduate level. 

It is probably fruitless to speculate on why a subject as basic, accessible, and important as
general relativity is not taught more widely as part of the undergraduate physics curriculum.
Limited time, limited resources, inertia, tradition, and misconceptions may all play a role.
Certainly it is not a lack of textbooks.  Refs.~\cite{Ber93}---\cite{Tou97}
 are a partial list of texts known to the
author\footnote{This list consists of texts known to the author,
published after 1975, and judged to be introductory.  It does not pretend 
to be either complete or selective, nor is it a representation that 
the texts are readily available.}
that treat general relativity in some way at an introductory level.

Available time is one of the obstacles to introducing general relativity at the undergraduate
level.  The deductive approach to teaching this subject (as for most others) is to assemble the
necessary mathematical tools, motivate the field equations, solve the equations in interesting
circumstances, and compare the predictions with observation and experiment.  This `math first' order
takes time to develop for general relativity which may not be available to either students or faculty.  This article
describes a different, `physics first' approach to introducing general relativity at the
junior/senior level.  Briefly, the simplest physically relevant solutions to the Einstein equation
are introduced {\sl first}, without derivation, as curved spacetimes whose properties and observable
consequences can be explored by a study of the motion of test particles and light rays.
This brings the student to interesting physical phenomena as quickly as possible.
It is the part of the subject most directly connected to classical mechanics and the part that
requires a minimum of `new' mathematical ideas.  Later the Einstein equation can be motivated and
solved to show where the solutions come from.  When time is limited this is a surer and more direct
route to getting at the applications of general relativity that are  important in contemporary
science. 

Section II expands very briefly on the importance of general relativity in contemporary physics.
Section III outlines the basic structure of the subject.  Sections IV
and V describe the `math first' and `physics first' approaches to 
introducing general relativity to undergraduate physics
majors.  This is not an even-handed comparison.   The `math first' approach is described only to contrast it with the
`physics first' approach which is advocated in this paper.
Section VI illustrates how ideas from classical mechanics can be used
to calculate important effects in general relativity. 
Section VII reports the personal experiences of the author in using the `physics first'
method.

\section{Where is General Relativity Important?}

Gravity is the weakest of the four fundamental forces at accessible energy scales.  The
ratio of the gravitational force to the electric force between two protons separated by a
distance $r$ is (in Gaussian electromagnetic units)
\begin{equation}
\frac{F_{\rm grav}}{F_{\rm elec}} = \frac{Gm^2_p/r^2}{e^2/r^2} =
\frac{Gm^2_p}{e^2} \sim 10^{-40}\, .
\label{twoone}
\end{equation}
Gravity might thus seem to be negligible.  But three other facts explain why it is
important and where it is important.  First, gravity is a universal force coupling to all
forms of mass and energy.  Second, gravity is a long-range force in contrast to the weak and
strong forces which are characterized by nucleus-size ranges and below.  Third, and most
importantly, gravity is unscreened.  There is no negative ``gravitational charge''; mass is
always positive.

These three facts explain why gravity is the dominant force governing the structure of the
universe on the largest scales of space and time --- the scales of astrophysics and
cosmology.  The strong and weak forces are short range.  The relatively much greater
strength of electromagnetic forces ensures that charges will be screened in an electrically
neutral universe like ours.  Only gravity is left to operate on very large scales.

Relativistic gravity --- general relativity --- is important for an object of mass $M$ and
size $R$ when
\begin{equation}
q \equiv  \frac{GM}{Rc^2} \sim 1\, .
\label{twotwo}
\end{equation}
Neutron stars $(q\sim .1)$ and black holes $(q\sim .5)$ are  relativistic objects
by this rough criterion.  So is our universe $(q \sim 1)$ if we take $R$ to be the present
Hubble distance and $M$ to be the mass within it. Figure 1 displays some phenomena
for which relativity is important and ones for which it is not.
\begin{figure}[t]
\centerline{\epsfxsize=3.3in \epsfbox{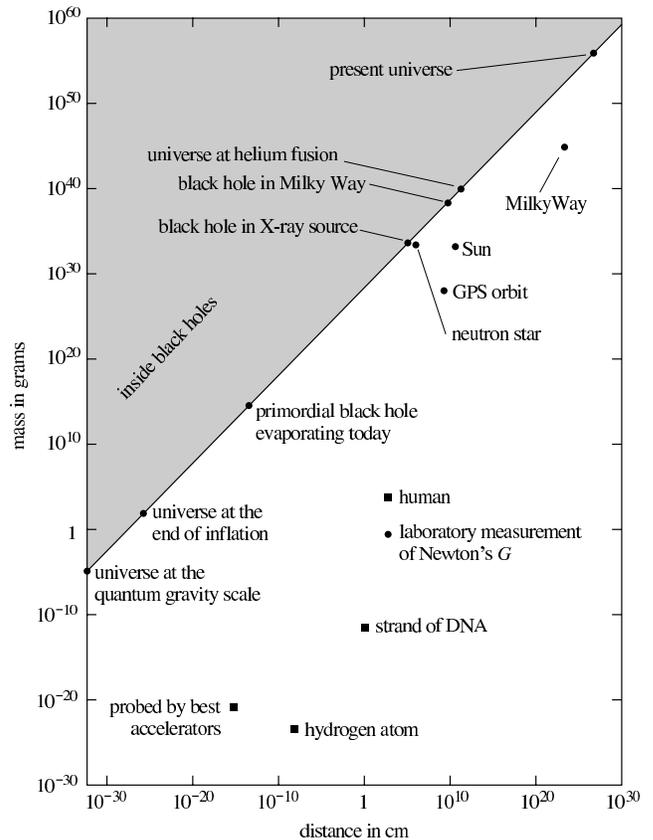}}
\caption{Where relativistic gravity is important. This figure shows selected phenomena plotted
on a graph of their characteristic mass $M$ {\it vs.}~their characteristic size $R$.  
The distance scale
ranges from the smallest considered in today's physics (the Planck scale) to the largest (the
size of the visible universe). Masses range from those of elementary particles to the mass in
the visible universe.  Phenomena above the line $2GM=c^2R$ are inside black holes and inaccessible.
Phenomena on that line, or close to it and below, are the ones for which general relativity is
important [{\it cf.}~\eqref{twotwo}] and are indicated by dots. Phenomena further away, for which
relativity is unimportant, are indicated by squares.  For the universe $R$ is the Hubble
distance and $M$ the mass inside it.  Our universe roughly evolves along the line $2GM=c^2R$ from the
smallest scales of quantum gravity to the largest characterizing present cosmology. (Reproduced
from \cite{Har03a} which was adapted from a figure prepared by C.~Will for 
\cite{cpg99}).  }
\end{figure}

General relativity can sometimes be important even when $q$ is small provided compensating
observational precision can be achieved. For the Sun $q\sim 10^{-6}$, yet the solar system is the domain
of the precision tests that confirm general relativity to as much as 1 part in $10^5$
\cite{Cassini}. For the Earth $q\sim 10^{-9}$, yet general relativistic effects are important for
the operation of the Global Positioning System (GPS) \cite{Ash02}. 

Relativistic gravity is also important on the smallest scales considered in contemporary
physics --- those of quantum gravity.  These are characterized by the Planck length $\ell$
\begin{equation}
\ell_{\rm Pl} = (G\hbar/c^3)^{\frac{1}{2}} \sim 10^{-33}\, {\rm cm}\, .
\label{twothree}
\end{equation}
This is much, much smaller than even the scale of the strong interactions $\sim 10^{-13}$ cm.
Yet, this is the scale which many contemporary explorers believe will 
naturally characterize the
final theory unifying the four fundamental forces including gravity.  This is the
characteristic scale of string theory.  This is the scale that will characterize the union
of the two great developments of twentieth century physics --- general relativity and
quantum mechanics.

The important point for this discussion is that the last few decades have seen dramatic growth
in observational data on the frontier of the very large, and an equally dramatic growth in
theoretical confidence in exploring the frontier of the very small.  Black holes, for example,
are no longer a theorist's dream.  They have been identified at the center of galaxies (including
our own) and in X-ray binaries.  They are central to the explanations of the most energetic
phenomena in the universe such as active galactic nuclei.  On even larger scales, it is now a
commonplace observation that cosmology has become a data driven science.  Cosmological
parameters once uncertain by orders of magnitude have been determined to accuracies of 10\%
\cite{Key, Wmap}. 

Adventures into Planck scale physics may be mostly in the minds of theorists, but the quest for
a unified theory of the fundamental forces including gravity is being pursued with impressive
vigor and confidence by a large community.  Indeed, at the big bang where large and
small are one, we should eventually see direct evidence of Planck scale physics.

For these reasons general relativity is increasingly central to today's physics.  It is
increasingly topical, increasingly connected with experiment and observation, and increasingly
integrated with other branches of physics\footnote{For more on the importance of general
relativity in contemporary physics, see {\it e.g.} \cite{cpg99}.}

\section{Key Ideas in General Relativity}

This section sketches a few key ideas in general relativity for those who may not be familiar
with the theory.  The intent is not to offer an exposition of these ideas.  That, after
all, is properly the task of texts on the subject.  Rather, the purpose is merely to mention ideas
that will occur in the subsequent discussion and to illustrate the simplicity of the conceptual
structure of the subject.

It's potentially misleading to summarize any subject in physics in terms of 
slogans.  However, the following three roughly stated ideas are central to
general relativity.

\begin{itemize}

\item {\sl Gravity is Geometry.} Phenomena familiarly seen as arising from gravitational forces
in a Newtonian context are more generally due to the curvature of geometry of four-dimensional
spacetime.

\item{\sl Mass-Energy is the Source of Spacetime Curvature.} Mass is the source of spacetime
curvature and, since general relativity incorporates special relativity, any form of energy is  also a source of spacetime curvature.

\item {\sl Free Mass Moves on Straight Paths in Curved Spacetime.}  In general relativity, the
Earth moves around the Sun in the orbit it does, not because of a gravitational force exerted by the Sun, but because it is following a straight path in the curved spacetime produced by the Sun.

\end{itemize}
Making these ideas more precise and more explicit is an objective of any course in general relativity.  We mention a few steps toward this objective here.

Points in four-dimensional spacetime can be located by four coordinates $x^\alpha, \alpha=0, 1,
2, 3$.  Coordinates are arbitrary provided they label points uniquely.  Generally, several
different coordinate patches are required to label all the points in spacetime.

The geometry of a spacetime is specified by giving the {\sl metric}, 
$g_{\alpha\beta} (x)$, where the $(x)$ indicates that the metric is
generally a function of all four coordinates.
The metric  determines the squared distance $ds^2$ in four-dimensional
spacetime between points separated by infinitesimal
coordinate intervals $dx^\alpha$. Specifically, 
\begin{equation}
ds^2=g_{\alpha\beta} (x) dx^\alpha dx^\beta
\label{threeone}
\end{equation} 
where a double sum over $\alpha$ and $\beta$ from 0 to 3 is implied. 
Integration of the $ds$ specfied by this
expression gives the distance along curves. 

Metrics satisfy the {\it Einstein equation}
\begin{equation}
R_{\alpha\beta} - \frac{1}{2}\ g_{\alpha\beta}\ R=\frac{8\pi G}{c^4}\ T_{\alpha\beta}
\label{threetwo}
\end{equation}
relating a measure of curvature on the left hand side 
to the energy-momentum tensor of matter on the
right.  This Einstein equation comprises 10 non-linear, partial differential equations for the
metric $g_{\alpha\beta}(x)$. An important example of a solution to the Einstein equation is the
Schwarzschild geometry giving the  metric in the empty space outside a spherically symmetric
black hole or star.  In standard Schwarzschild spherical coordinates $x^\alpha=(t, r, \theta, \phi)$
this is
\begin{align}
ds^2  = & - \left(1-\frac{2GM}{c^2r}\right)\, (cdt)^2  +  
\left(1-\frac{2GM}{c^2r}\right)^{-1} dr^2 \nonumber 
\\
  &+ r^2 \left(d\theta^2+\sin^2 \theta d\phi^2\right)
\label{threethree}
\end{align}
where $M$ is the mass of the black hole or star.  This, to an excellent approximation, describes
the curved spacetime outside our Sun. 

Test particles with masses too small to affect the ambient geometry move on straight paths in
it.  More precisely, they move between any two points, $A$ and $B$, in spacetime on a world line (curve)  of
stationary proper time $\tau$. The proper time along a  world line is the distance along it measured in time units.  Thus, $d\tau^2 = - ds^2/c^2$.  (The negative sign is so $d\tau^2$ is positive along the
world lines of particles which always move with less than the speed of light). Curves of
stationary proper time are called {\it geodesics}.
%\begin{equation}
%\delta \int^B_A d\tau=0\, .
%\label{threefour}
%\end{equation}

The world line of a particle through spacetime from $A$ to $B$ can be
described by giving its  
coordinates $x^\alpha(\lambda)$ as a function of any parameter that takes fixed values on the end points.   For instance
using the Schwarzschild metric \eqref{threethree}, the principle of stationary 
proper time takes the form:
\begin{align}\delta \int^B_A & d\tau
=\delta\int^B_A d\lambda\, \frac{1}{c^2} \biggl[\left(1-\frac{2GM}{c^2r}\right)
(c\dot t)^2 \nonumber \\ & - \left(1-\frac{2GM}{c^2r}\right)^{-1} {\dot r}^2   - r^2 \left(\dot \theta^2+
\sin^2\theta\, \dot\phi^2\right)\biggr]^{\frac{1}{2}}=0
\label{threefive}
\end{align}
where a dot denotes a derivative with respect to $\lambda$ and $\delta$
means the first variation as in classical mechanics.
%\delta\int^B_A d\lambda\, \frac{1}{c^2} & \Biggl[\left(1-\frac{2GM}{c^2r}\right)
%(c\dot t)^2 - \left(1-\frac{2GM}{c^2r}\right)^{-1}\\\nonumber
%& \dot r^2 - r^2 \left(\dot \theta^2+
%\sin^2\theta\, \dot\phi^2\right)\Biggr]^{\frac{1}{2}}=0

The variational principle \eqref{threefive} for stationary proper time has the same form as the
variational principle for stationary action in classical mechanics.  The Lagrangian $L(\dot
x^\alpha, x^\alpha)$ is the integrand of \eqref{threefive}.  Lagrange's equations are the
geodesic equations of motion.  Their form can be made especially simple by choosing proper time
$\tau$ for the parameter $\lambda$.  From them one can deduce the conservation of energy, the
conservation of angular momentum, and an effective equation for radial motion in the
Schwarzschild geometry as we will describe in Section VI
[cf.\eqref{sixone}].  With that, one can calculate a spectrum of phenomena
ranging from the precession of planetary orbits to the collapse to a black hole.

When extended to light rays and implemented in  appropriate metrics, the geodesic equations
are enough to explore most of the  important applications of relativistic gravity displayed in Table 1.

\begin{table*}
\begin{tabular}{|llllll|}
\multicolumn{6}{c}{\bfseries Table 1}\\
\multicolumn{6}{c}{\bfseries Some Important Applications of General Relativity}\\\hline
Global Positioning System (GPS) &&&&& Lense-Thirring precession of a gyroscope\\
Gravitational redshift &&&&& Cosmological redshift\\
Bending of light by the Sun &&&&& Expansion of the universe\\
Precession of Mercury's perihelion &&&&& Big-bang\\
Shapiro time delay &&&&& Cosmic Background Radiation\\
Gravitational lensing &&&&& The fate of the universe\\
Accretion disks around compact objects &&&&& Propagation of gravitational waves\\
Determining parameters of binary pulsars &&&&& Operation of gravitational wave detectors\\
Spherical gravitational collapse &&&&& X-ray sources\\
Formation of black holes &&&&& Active Galactic Nuclei\\
Hawking radiation from black holes &&&&& Neutron stars\\
Frame-dragging by a rotating body &&&&&  \\\hline

\end{tabular}
\end{table*}

\section{Teaching General Relativity --- Math First}

The deductive approach to teaching many subjects in physics is to

\begin{enumerate}

\item Introduce the necessary mathematical tools;
\item Motivate and explain the basic field equations;
\item Solve the field equations in interesting circumstances;
\item Apply the solutions to make predictions and compare
with observation and experiment.

\end{enumerate}

For electromagnetism, (1)--(4) are, {\it e.g.} (1) Vector calculus, (2) Maxwell's equations, (3)
boundary value problems, the fields of point particles, radiation fields, etc., (4) charged
particle motion, circuits, wave guides and cavities, antennas, dielectric and magnetic
materials, magnetohydrodynamics --- a list that could very easily be extended.  For
gravitation (1)--(4) are {\it e.g.} (1) differential geometry, (2) the Einstein equation and the
geodesic equation, (3) the solutions for spherical symmetry, cosmological models, gravitational
waves, relativistic stars, etc. Table 1 lists some of the applications of general relativity that
constitute (4).

This deductive order of presentation is logical; it is the order used by the great classic texts 
\cite{LL62, MTW70, Wei72, Wal84};
and it is the order used in standard graduate courses introducing the subject at an advanced
level. But the deductive order does have some drawbacks for an elementary introduction to physics
majors in a limited time.

Differential geometry is a deep and beautiful mathematical subject. However, even an elementary
introduction to its basic ideas and methods are not a part of the typical advanced calculus tool
kit acquired by physics majors in their first few years.  This is `new math'.  
In contrast,  
the vector calculus central to electromagnetism  {\it is} part of this tool kit.

It is possible at the undergraduate level to give an introduction to the basic mathematical ideas
of manifolds, vectors, dual vectors, tensors, metric, covariant derivative, and curvature.  Indeed,
many students feel empowered by learning new mathematics. But
it does take time.  It also must be practiced. The author's experience is that many 
students at this
level need considerable exercise before they are able to accurately and efficiently manipulate
tensorial expressions and feel at home with the four-dimensional mathematical concepts necessary
to formulate Einstein's equation.  When time is limited, pursuing the deductive order may leave
little available for the interesting applications of general relativity.

Further, solving the Einstein equation to exhibit physically relevant spacetime geometries is a
difficult matter. Their non-linear nature means that there is no known general solution outside
of linearized gravity.  Each new situation, {\it e.g.} spherical symmetry, homogeneous and
isotropic
cosmological models, gravitational plane waves, is typically a new problem in applied
mathematics.  Deriving the solutions only adds to the time expended before 
interesting
applications can be discussed.

Many of the successful introductory texts in general relativity 
follow this logical order
at various levels of compromise.   In the author's opinion, an outstanding example
is Bernard Schutz's classic {\sl A First Course in General Relativity} \cite{Sch85}.
In the next section we consider a different way of introducing general relativity to
undergraduates.

\section{Teaching General Relativity --- Physics First}

Electricity and magnetism are not usually presented in introductory (freshman) courses in the
deductive order described in the previous section. Specifically, we do not usually
first develop vector calculus, then exhibit Maxwell's equations, then solve for the fields of
charges, currents, and radiation, and finally apply these to realistic electromagnetic
phenomena.  Rather, the typical course posits the fields of the simplest physically relevant
examples, for instance the electric field of a point charge, the magnetic field of a straight
wire, and the electromagnetic plane wave.  These are used to build understanding of fields and
their interaction with charges for immediate application to demonstrable electromagnetic
phenomena.  Maxwell's ten partial differential equations and their associated gauge and
Lorentz invariances are better appreciated later, usually in a more advanced course.
 
General relativity can be efficiently introduced at an  intermediate (junior/senior) level
following the same `physics first' model used in introducing electromagnetism.  Specifically:

\begin{enumerate}

\item Exhibit the simplest physically important spacetime geometries {\it first}, without
derivation;

\item Derive the predictions of these geometries for observation by a study of the orbits of
test particles and light rays moving in them;

\item Apply these predictions to realistic astrophysical situations and compare with
experiment and observation;

\item Motivate the Einstein equation and solve it to show where the
spacetime geometries posited in (1) come from.

\end{enumerate}
These are essentially the same four elements that comprise the deductive approach described in
the previous section, but in a different order.  That order has considerable advantages for
introducing general relativity at an intermediate level as we now describe:

\subsection{Indications}

\noindent {\sl Less `New Math' Up Front:}
To exhibit a spacetime geometry, the only `new math' ideas required are the metric and its 
relation to distances in space and time.  To analyze the motion of test particles in these
geometries, only the notions of four-vectors and geodesics are needed.  These three new
mathematical ideas are enough to explain in detail a wide range of physical phenomena, such as most of those in  Table 1.
Further, these three new mathematical ideas are among the simplest parts of a
relativist's tool kit to introduce at an intermediate level.  Four-vectors are often familiar
from special relativity.  Geodesics viewed as curves of extremal proper time 
are special cases of
Lagrangian mechanics.  The idea of a metric can be motivated from the theory of surfaces in
three-dimensional flat space.  A general theory of tensors as linear maps from vectors into
the real numbers is not required because only one tensor --- the metric --- is ever used.
\vskip .13 in
\noindent {\sl The Simplest Spacetimes are the Most Physically Relevant:}

\begin{itemize}

\item The Sun is approximately spherical.
\item Spherical black holes exhibit many characteristic properties of the most general black
hole.
\item The universe is approximately homogeneous and isotropic on scales above 100 megaparsecs.
\item Detectable gravitational waves are weak.

\end{itemize}
These four facts mean that the simplest solutions of the Einstein equation are the ones
most relevant for experiment and observation.  The static, spherically 
symmetric Schwarzschild geometry \eqref{threethree} describes the solar system experimental tests, spherically symmetric gravitational
collapse, and spherical black holes.
The exactly homogeneous and isotropic Friedman-Robertson-Walker (FRW) models 
provide an
excellent approximation to the structure and evolution of our universe from the big bang to
the distant future. The linearized solutions of the Einstein equation about a flat space background
describe detectable gravitational waves. A `physics first' treatment of 
general relativity that concentrates on the simplest solutions of the 
Einstein equation is thus immediately relevant for physically realistic
and important situations.  
\vskip .13 in
\noindent{\sl No Stopping before Some Physics:} Students with different levels of experience, preparation, abilities, and
preconceptions will take different lengths of time to acquire the basic concepts of general
relativity.  Beginning with the applications guarantees that, wherever the course ends, students
will have gained some understanding of the basic physical phenomena (Table 1) which make general
relativity so important and not simply of a mathematical structure which is the prerequisite for a
deductive approach.
\vskip .13in
\noindent{\sl More Concrete, Less Abstract:} Beginning with the
applications rather than the abstract structure of the theory is easier
for some students because it is more concrete. Beginning with the 
applications is also a surer way of driving home that general relativity
is a part of physics whose predictions can be observationally tested and
not a branch of mathematics.  
\vskip .13 in
\noindent{\sl Fewer Compromises:} The analysis of the motion of test particles and
light rays in the simplest geometries can be carried out in essentially the same way as it is done in advanced textbooks.
No compromises of method or generality are needed. 
\vskip .13 in
\noindent{\sl Flexibility in Emphasis:}  Beginning with applications allows 
enough time to construct courses with different emphases; for instance on 
black holes, gravitational waves, cosmology, or experimental tests.
\vskip .13 in
\noindent{\sl Closer to the Rest of the Undergraduate Physics
Curriculum:} Calculations of  the orbits of test particles and light rays to explore  curved spacetimes are exercises in mechanics.  
The symmetries of the simplest important solutions imply conservation laws.
These can be used to reduce the calculation of 
orbits to one-dimensional motion in effective potentials.  Even the
content of the Einstein equation can be put in this form for simple situations.  This allows the intuition and techniques developed in intermediate mechanics to be
brought to bear,  both for  qualitative understanding and quantitative prediction.  Conversely, this kind of example serves to extend and reinforce an understanding of mechanics.  Indeed, in the author's experience a few
students are surprised to find that mechanics is actually useful for
something. The examples discussed in the next section will help
illustrate the close connection between calculating geodesics and
undergraduate mechanics.  
\vskip .13in 
\noindent{\sl Fewer Prerequisites:}  The close connection with mechanics described above and in
the next section means that
the only essential prerequisite to a `physics first' exposition of general relativity is an
intermediate course in mechanics.  Neither quantum mechanics nor 
electrodynamics are necessary.%
%\footnote{An exposition of the sources of gravitational waves in linearized 
%gravity {\it can} benefit from the close
%analogy to the production of electromagnetic radiation.  However, that 
%connection is not essential and neither need the production of gravitational 
%radiation be a part of every introductory course.}
Some acquaintance with special relativity is useful, but its brief
treatment in many first year courses means that it is usually necessary to
develop it {\it de novo} at the beginning of an introductory course in
general relativity. Intermediate mechanics is thus the single essential 
physics prerequisite. This means that an introductory `physics first'
course in general relativity can be accessible to a wider range of physics 
majors at an earlier stage than courses designed to introduce
students to other frontier areas.
\vskip .13 in
\noindent{\sl Closer to Research Frontiers:}  Beginning with the applications means that students are closer sooner to the contemporary frontiers of astrophysics and particle physics that they can hear about in the seminars, read about in the newspapers, and see on popular television programs.  
\vskip .13in 
\noindent{\sl More Opportunities for Undergraduate Participation in Research:}  The
applications  of general relativity provide a broad range of topics for students to pursue independent study
or even to make research contributions of contemporary interest. More importantly,  it is possible to identify problems from the applications that are conceptually and technically
accessible to undergraduate physics majors and can be completed in the limited time
frame typically available. Problems that involve solving for the behavior of test particles,
light rays, and gyroscopes are examples, as are questions involving linear gravitational
waves, black holes, and simple cosmological models. The `physics first' approach to
teaching general relativity enables undergraduate participation in research because it
treats such applications first.  
\vskip .13 in
\noindent{\sl Specialized Faculty Not Needed:}  Learning a subject
while teaching it, or learning it better, is a part of every physics instructor's experience.  The
absence of previous instruction means that teaching an introductory
general relativity course will often be
the first exposure to the subject for many faculty.  The process of learning by faculty is made
easier by a `physics first' approach for the same reasons it is easier for students.  A wider variety
of faculty will find this approach both familiar and manageable.  Specialists in gravitational
physics are not necessary.

\subsection{Counterindications}

No approach to teaching is without its price and the `physics first' approach to introducing
general relativity is no exception.  The obvious disadvantage is that it does not follow the
logically appealing deductive order, although it can get to the same point in the end.  
A `physics
first' approach to introducing general relativity {\it may} not be indicated 
in at least two
circumstances:  First, when there is enough time in the curriculum and enough 
student commitment to pursue
the deductive order;  second, when students already have significant 
preparation in differential geometry. Even then however it is a possible
alternative. A `physics first' approach is 
probably not indicated when the mathematics is of central interest, as for 
students concentrating in mathematics, and for physics students who study 
Einstein's theory mainly as an introduction to the
mathematics of string theory.

\section{Particle Orbits Outside a Spherical Star or Black Hole}

This section illustrates how the effective potential method developed in typical undergraduate mechanics courses can be applied to important problems in
general relativity. Of the several possible illustrations, just one is 
considered here --- the relativistic effects on test 
particle orbits outside a spherical star or black hole. Applications of this include the
precession of perihelia of planets  in the solar system and the location of the innermost stable circular orbit (ISCO) in an accretion disk powering an X-ray source. 
These examples also serve to illustrate how near the calculations of such effects are
to starting principles in a `physics first' approach to general relativity. 

Nothing  more than sketches of the calculations are intended. 
For more detail and the precise meaning of
any quantities involved, the reader should consult any standard text\footnote{We follow, with one minor simplifying exception,  the notation in \cite{Har03a}.}.

The Schwarzschild geometry specified in \eqref{threethree} describes the 
curved spacetime outside a static, spherically symmetric body of  
mass $M$. This is the geometry outside the Sun to an excellent
approximation,  and is the
geometry outside a spherical black hole. The motion of test particles is specified by
the principle of stationary proper time in \eqref{threefive}. The argument in 
the square bracket of that equation can be thought of as a Lagrangian 
$L(\dot t, \dot r, \dot\theta, \dot\phi, r)$. Lagrange's equations are
the geodesic equations. 

Time translation invariance implies a conserved energy per unit rest mass 
$\cal E$ related to $\partial L/\partial{\dot t}$ .  Spherical symmetry
implies a conserved angular momentum per unit rest 
mass \footnote{In \cite{Har03a}, $\cal E$ and $\ell$ are defined as an energy 
and angular
momentum  per unit rest energy rather than per unit rest mass as here.
With that definition of $\cal E$ both terms on the right hand side of 
\eqref{sixone a} would be divided by $c^2$. That is the minor exception alluded to in the previous footnote.} 
$\ell$ for
orbits in the equatorial plane which is proportional to $\partial
L/\partial{\dot\phi}$.
A third integral\footnote{`Integral' here is used
in the sense of classical mechanics, not in the sense of the inverse of
differentiation.} of the motion $L=-1.$ arises just from the definition 
of proper time.  This integral can be combined with the other to to find an 
effective energy integral for the radial motion: 
\begin{subequations}
\label{sixone}
\begin{equation}
{\cal E} = \frac{1}{2}\ \left(\frac{dr}{d\tau}\right)^2 + V_{\rm eff} (r)
\label{sixone a}
\end{equation}
where
\begin{equation}
V_{\rm eff} (r) = - \frac{GM}{r} + \frac{\ell^2}{2r^2} - \frac{GM\ell^2}{c^2r^3}\, .
\label{sixone b}
\end{equation}
\end{subequations}%
Here, $r$ is the Schwarzschild radial coordinate of the test particle 
and $\tau$ 
\begin{figure}[t]
\centerline{\epsfxsize=3in \epsfbox{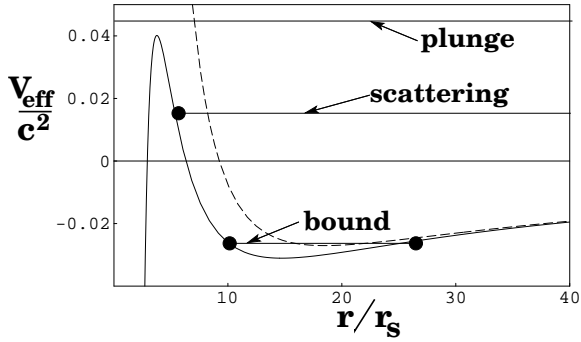}}
\caption{The effective potential $V_{\rm eff}(r)$ for the radial motion of particles
in the Schwarzschild geometry outside a spherical star or black hole.
The solid line shows $V_{\rm eff}(r)/c^2$ defined in
\eqref{sixone b} plotted against $r/r_s$ where $r_s=GM/c^2$.  
The value
of the angular momentum parameter is $\ell=4.3GM/c^2$. The dotted line shows the
Newtonian potential (the first two terms in \eqref{sixone b}) for the same value of $\ell$. The
qualitative behavior of the orbits depends on the value of the energy parameter ${\cal E}$.
Values are shown for plunge orbits, scattering orbits, and bound orbits such as those executed by
the planets in their motion about the Sun. }
\end{figure}
is the proper time along
its world line.  The mass of the test particle is absent from  these expressions.  
It cancels out because of the
equality of gravitational and inertial mass.
Eq. \eqref{sixone a} has the same form as the energy integral for a
Newtonian central force problem. 
The first two terms in the effective potential \eqref{sixone b} have the same 
form as a Newtonian gravitational potential with a Newtonian centrifugal 
barrier.  The third term provides a general relativistic correction to 
the Newtonian effective potential. Figure 2 shows its effects. 

Circular orbits illustrate the importance of these effects. 
Newtonian gravity permits only one stable circular orbit
for each $\ell$. But in general relativity there are {\it two} circular
orbits for values of $\ell$ such at that used in Figure 2.  There is a
stable circular orbit at the {\it minimum} of the effective potential such as one
approximating the orbit of the Earth in its progress around the Sun. 
In addition there is an unstable circular orbit at the radius of the
{\it maximum} of the effective potential. 

The radii of the stable circular orbits are easily found from
\eqref{sixone b}:
\begin{equation}
r_{\rm stab.circ.} = \frac{\ell^2}{2GM}\bigg\{1+\bigg[1
-12\left(\frac{GM}{c\ell}\right)^2\bigg]^{\frac{1}{2}} \bigg\} .
\label{circ}
\end{equation}
For sufficiently small $\ell$ there are {\it no} stable circular orbits.
That is because the effective potential \eqref{sixone b} is everywhere
attractive for low $\ell$. In contrast to Newtonian physics, general relativity therefore implies that
there is an {\it innermost {\rm (smallest
$r$)} stable circular orbit} (ISCO) whose radius is 
\begin{equation}
r_{\rm ISCO} =6GM/c^2 
\label{isco}
\end{equation} 
which is $1.5$ times the characteristic radius of the black hole
$r_s=2GM/c^2$. 

The ISCO is important for the astrophysics of black holes. 
The spectra of X-ray sources exhibit lines whose observed shapes can in
principle be used to infer the properties of the black hole engine
\cite{Ironlines}. The shape of the line is determined by several factors 
but importantly affected by the gravitational redshift. That is maximum
for radiation from parts of the accretion disk closest
to the black hole, ie from the ISCO.  This defines the red end of the 
observed line\footnote{Realistic black holes are generally rotating, but
the analysis then is not qualitatively different from that for  the non-rotating
Schwarzschild black hole.}.

Another important prediction of general relativity derivable from the 
effective potential is the shape of bound orbits such as those of the
planets. The shape of an orbit in the equatorial plane ($\theta=\pi/2$)
may be specified by giving the azimuthal angle $\phi$ as a function of
$r$. The orbits close if the total angle $\Delta\phi$ swept out in the
passage away from the inner turning point and back again is $2\pi$. This can 
be calculated from  \eqref{sixone} and the angular momentum 
integral $\ell=r^2(d\phi/d\tau)$. Writing $d\phi/dr=(d\phi/d\tau)/(dr/d\tau)$ 
and using these two relations gives an expression for $d\phi/dr$ as a function 
of $r$, ${\cal E}$, and $\ell$ which can be integrated. The result is
\begin{equation}
%\Delta\phi = 2 \int dr \frac{\ell}{r^2}\frac{1}{\left[2\left({\cal E}
%-V_{\rm eff}(r)\right)\right]^{1/2}} 
\Delta\phi = 2 \int dr (\ell/r^2)\left[2\left({\cal E}
-V_{\rm eff}(r)\right)\right]^{-1/2} 
\label{prec}
\end{equation} 
where the integral is from the radius of the inner turning point to the
outer one.  When the relativistic term in \eqref{sixone b} is absent,
$\Delta\phi=2\pi$ for all $\cal E$ and $\ell$. That is the closing of
the Keplerian ellipses of Newtonian mechanics. The relativistic
correction to $V_{\rm eff}$ makes the orbit precess by small amount 
$\delta\phi =
\Delta\phi -2\pi$ on each pass. To lowest order in $1/c^2$ this is
\begin{equation}
\delta\phi = 6\pi \left(\frac{GM}{c\ell}\right)^2 \ .
\label{prec1}
\end{equation} 
In the solar system the precession is largest for Mercury but still only
$43''$ per century. The confirmation of that prediction \cite{Sha90} is
an important test of general relativity. 

The purpose of this section was not to explain or even emphasize the two 
effects of general relativity on particle orbits that were described
here. Rather it was to show three things.
First, that a standard technique developed in undergraduate mechanics
can be used to calculate important predictions of general relativity. 
The second purpose was to show how close these important applications can come
to starting principles in a `physics first' approach to teaching general
relativity. Introduce the Schwarzschild geometry \eqref{threethree}, use
the principle of stationary proper time \eqref{threefive} to find the 
geodesic equations or their integrals \eqref{sixone}, use the effective 
potential method to qualitatively and quantitatively understand important 
properties of the orbits e.g. \eqref{isco} and \eqref{prec}. That is
just three steps from the basic ideas of metric and geodesics to
important applications. Third,  both of the applications treated here
can be immediately related to contemporary observation and experiment.   

Particle orbits in the Schwarzschild geometry are not the only important
problems in which the effective potential method is useful. Motion in
the geometry of a rotating black hole, the motion of test light rays in
the Schwarzschild geometry, and the evolution of the Friedman-Robertson-Walker 
cosmological models are further examples where it can be usefully
applied. 

\section{Conclusion}

A one quarter ($\sim$ 28 lectures) `physics first' course in general relativity has been a
standard junior/senior elective for physics majors at the University of California, Santa
Barbara for approximately thirty years.  In the limited span  of a quarter the author is usually able to review special relativity, motivate 
gravity  as geometry, derive the orbits in the Schwarzschild geometry in 
detail, describe the experimental tests, introduce black holes, and develop 
the Friedman-Robertson-Walker cosmological models. A semester provides more 
opportunities. 

At Santa Barbara this course is routinely taught by faculty from
many different areas of physics ---  general relativity of course, but also 
elementary particle physics and astrophysics. It has been taught by
both theorists and experimentalists.  For some of these colleagues 
teaching this course was their first serious experience with general 
relativity. They usually report that they were successful and enjoyed
it.  

In the author's experience students are excited by general relativity
and motivated to pursue it. Often it is their
first experience with a subject directly relevant to current research. 
It is one of the few contemporary subjects that can be taught without
quantum mechanics or electromagnetism.

The author has written a text based on the `physics first' approach \cite{Har03a} which comes with a solutions manual available to instructors for the approximately 400 problems of graded levels of difficulty.  

`Physics first' is  not the only way of introducing general relativity to 
undergraduate physics majors, but it works.

\acknowledgments 
Thanks are due to Bill Hiscock, Ted Jacobson,  Richard Price, 
and Francesc Roig for critical readings
of the manuscript. 
Preparation of this paper was supported in part by NSF grant PHY02-44764.

\end{document}